\documentclass[12pt,a4paper]{article}
\usepackage{amsfonts,latexsym}
\usepackage{amsmath,amssymb}
\usepackage{graphicx,color}

\renewcommand{\thefootnote}{\fnsymbol{footnote}}

\begin{document}

\vspace{12mm}
\begin{center}
{{{\Large {\bf Scalarizations of qOS-extremal black hole \\and Aretakis instability  }}}}\\[10mm]
{Yun Soo Myung\footnote{e-mail address: ysmyung@inje.ac.kr}}\\[8mm]

{Center for Quantum Spacetime, Sogang University, Seoul 04107, Republic of  Korea\\[0pt] }

\vspace{12mm}

\end{center}
 \begin{abstract}
    \noindent We study  scalarization of quantum Oppenheimer-Snyder (qOS)-extremal black hole in the Einstein-Gauss-Bonnet-scalar theory.
    This black hole is described by mass ($M=4\sqrt{\alpha}/3\sqrt{3}$) with $\alpha$ quantum parameter. From studying onset of scalarization, we find the appearance of the single branch of scalarized qOS-extremal black holes. To obtain the tachyonic scalar cloud being  a seed to generate the single branch of scalarized qOS-extremal black holes, we consider the near-horizon geometry of the Bertotti-Bobinson (BR) spacetime. In this case, we find that  the appearance of a large  scalar cloud at the horizon ($\rho=0$) is a new feature to represent  onset scalarization of extremal black holes for tachyon with negative mass, but it is not related to the Aretakis instability of a propagating scalar with standard mass around the BR spacetime, showing polynomial instability of the ingoing time $v$. The Aretakis instability is connected  to the  scalar cloud  with standard mass, indicating the blow-up  at the horizon.
\end{abstract}

\vspace{1.5cm}

\hspace{11.5cm}{Typeset Using \LaTeX}
\newpage
\renewcommand{\thefootnote}{\arabic{footnote}}
\setcounter{footnote}{0}

\vspace{2mm}

\section{Introduction}

The quantum Oppenheimer-Snyder (qOS)-black hole was recently found from investigating the qOS gravitational collapse within the loop quantum cosmology~\cite{Lewandowski:2022zce}.
However, one does not know  its action $\mathcal{L}_{\rm qOS}$ to give the qOS-black hole described by mass ($M$) and quantum parameter ($\alpha$) as a direct solution.
Various studies of this black hole including  quasinormal mode analysis for tensor and scalar perturbations~\cite{Skvortsova:2024msa}, thermodynamics~\cite{Dong:2024hod,Mazharimousavi:2025lld,Panotopoulos:2025ygq}, shadow radius~\cite{Yang:2022btw,Ye:2023qks}, and  scalarization within the Einstein-Gauss-Bonnet-scalar (EGBS) theory~\cite{Chen:2025wze,Myung:2025pmx} were explored.

On the other hand, extremal black holes have  played an important role in various aspects. They possess
zero Hawking temperature and zero heat capacity  and thus, are expected to bring us valuable insights into the black hole
thermodynamics~\cite{Charmousis:2010zz} and the Hawking radiation~\cite{Angheben:2005rm}.
In the astrophysics, it was proposed that many astrophysical black holes are nearly extremal~\cite{Volonteri:2004cf,Gou:2013dna}.
To understand  the nature of the extremal black holes, it is valuable  to investigate the dynamical properties
of test fields and particles propagating around them. In this direction, Aretakis~\cite{Aretakis:2011ha} has discussed late-time behaviors of  massless scalars  in the
extremal Reissner-Nordstr\"om  black holes, leading to that  higher-order transverse
derivatives of the scalar fields blow up polynomially in the ingoing time $v$ on the event horizon.  This blow-up on the horizon is called the Aretakis
instability. In the  near-horizon approximation, the leading
behavior of a  massive scalar field was  described by power-law tails, showing the
Aretakis instability too~\cite{Katagiri:2021xig,Chen:2025sim}.

The no-hair theorem implies  that a black hole can be completely described  by  three externally observable parameters: mass ($M$), electric charge ($Q$), and rotation parameter ($a$) in Einstein-Maxwell gravity
\cite{Carter:1971zc,Ruffini:1971bza}. If a scalar field is minimally coupled to  gravitational and electromagnetic  fields, there is no scalar hair~\cite{Herdeiro:2015waa}.
However, its evasion  occurred in the context of scalar-tensor theories possessing the nonminimal scalar coupling to either Gauss-Bonnet (GB) term~\cite{Doneva:2017bvd,Silva:2017uqg,Antoniou:2017acq} or to Maxwell term~\cite{Herdeiro:2018wub,Myung:2018vug}, where the former is called GB$^+$ scalarization triggered  from tachyonic instability with a positive coupling parameter.

Furthermore, the  spin-induced (GB$^-$) scalarization
of Kerr black holes with rotation parameter $a$ was demonstrated   for $a_c(=0.5)<a\le 1$ in the EGBS
theory with a negative  coupling parameter~\cite{Cunha:2019dwb,Collodel:2019kkx,Dima:2020yac,Herdeiro:2020wei,Berti:2020kgk}. In this direction, we would like to mention that GB$^-$ scalarization was realized for a very narrow region of  $q_c(=0.9571)<q\le 1$ with $q=Q/M$ in the Einstein-Gauss-Bonnet-Maxwell-scalar (EGBMS) theory~\cite{Brihaye:2019kvj,Herdeiro:2021vjo,Hod:2023nyt}. Here, the charge $Q$ played a role of the rotation parameter $a$.
Also,  we note that the quantum parameter ($\alpha$)-mass ($M$) induced scalarization (GB$^-$) was studied in the EGBS theory with the unknown action $\mathcal{L}_{\rm qOS}$~\cite{Chen:2025wze,Myung:2025pmx}, showing the allowed region for quantum parameter as $\alpha_c(=1.2835)<\alpha\le \alpha_e(=1.6875)$ with $M=1$. The mass allowed region for GB$^-$ scalarization is given by a narrow region of  $M_{\rm rem}(=0.7698)<M\le M_c(=0.8827)$ with $\alpha=1$.
Furthermore, it was shown that  two branches of positive ($\gamma>0$) and negative ($\gamma<0$) coupling parameter  are available  in the  spontaneous scalarization of charged  black holes at the approach to extremality  in the EGBMS theory~\cite{Brihaye:2019kvj}.  These authors claimed that  the presence of  negative branch is related to  the near-horizon geometry of AdS$_2\times S^2$.

In the present work, we wish to investigate  scalarization of qOS-extremal black holes described by mass ($M$) in the EGBS theory with the coupling parameter  $\lambda$.
In this case, the mass $M$  is regarded as the main parameter, whereas the quantum parameter $\alpha$  is redundant because of the relation $\alpha=27M^2/16$.
Also, it is interesting to note that  its temperature and heat capacity are always  zero and critical onset parameter $M_c$ disappears, implying the absence  of the upper bound on the mass $M$.

Studying on the onset of GB$^-$ scalarization with $\lambda<0$, we find  the sufficiently unstable region of $0<M\le M_{sEEH}(=0.96\sqrt{-\lambda})$, which predicts the appearance of the single branch of scalarized qOS-extremal black holes. This is  similar to the sufficiently unstable region of $0<M\le M_{S}(=1.1\sqrt{\lambda})$ for GB$^+$ scalarization of Schwarzschild black holes with $\lambda>0$, but it accommodates infinite branches of scalarized black holes.  However, we could not obtain its tachyonic scalar cloud which may be  a seed to generate the single branch of scalarized qOS-extremal black holes.
This is because numerical methods cannot solve the Klein-Gordon equation to find out scalar clouds in the extremal black hole background~\cite{Richartz:2015saa,Senjaya:2025pyv}.
This forces the numerical investigation to end at the near-extremal limit~\cite{Hod:2017gvn}.

To obtain the tachyonic scalar cloud with  the tachyonic mass $\mu^2=8\lambda<0$, we introduce  the near-horizon geometry of the Bertotti-Bobinson (AdS$_2\times S^2$) spacetime.
In this case, we find  the appearance of the large  scalar cloud at the horizon ($\rho=0$) which  is a new feature to represent  onset of scalarization for extremal black holes.
However, this  is not related to the Aretakis instability of a propagating scalar with standard mass $\mu^2=8\lambda>0$  around the AdS$_2\times S^2$ spacetime,  which indicates  polynomial instability of the ingoing time $v$. This instability  is  connected  to the  scalar cloud  with standard mass, indicating the blow-up  at the horizon.

\section{qOS-extremal black hole}
The EGBS theory with the unknown qOS action ${\cal L}_{\rm qOS}$ takes the form~\cite{Chen:2025wze,Myung:2025pmx}    as
\begin{equation}
\mathcal{L}_{\rm EGBSq}=\frac{1}{16 \pi}\Big[ R-2\partial_\mu \phi \partial^\mu \phi+\lambda f(\phi) {\cal R}^2_{\rm GB}+{\cal L}_{\rm qOS}\Big],\label{Action1}
\end{equation}
where $\phi$ is the scalar field,  a quadratic coupling function $f(\phi)=2\phi^2$, and  $\lambda$ is a coupling constant with length dimension 2. Also, ${\cal R}^2_{\rm GB}=R^2-4R_{\mu\nu}R^{\mu\nu}+R_{\mu\nu\rho\sigma}R^{\mu\nu\rho\sigma}$ denotes the GB term.

The Einstein  equation  with  $G_{\mu\nu}=R_{\mu\nu}-(R/2)g_{\mu\nu}$ is derived as
\begin{eqnarray}
 G_{\mu\nu}=2\partial _\mu \phi\partial _\nu \phi -(\partial \phi)^2g_{\mu\nu}+\Gamma_{\mu\nu}+T^{\rm qOS}_{\mu\nu}, \label{equa1}
\end{eqnarray}
where $\Gamma_{\mu\nu}$   is given by
\begin{eqnarray}
\Gamma_{\mu\nu}&=&2R\nabla_{(\mu} \Psi_{\nu)}+4\nabla^\alpha \Psi_\alpha G_{\mu\nu}- 8R_{(\mu|\alpha|}\nabla^\alpha \Psi_{\nu)} \nonumber \\
&+&4 R^{\alpha\beta}\nabla_\alpha\Psi_\beta g_{\mu\nu}
-4R^{\beta}_{~\mu\alpha\nu}\nabla^\alpha\Psi_\beta  \label{equa2}
\end{eqnarray}
with
\begin{equation}
\Psi_{\mu}=\lambda f'(\phi)\partial_\mu \phi.
\end{equation}
Here, its energy-momentum tensor may take the form
\begin{equation}\label{q-em}
T^{\rm qOS, \nu}_{\mu}=\frac{3\alpha M^2}{r^6}{\rm diag}[-1,-1,2,2],
\end{equation}
 which corresponds to the anisotropic energy-momentum tensor.
 An alternative   action  for $\mathcal{L}_{\rm qOS}$ was   suggested  by a nonlinear electrodynamics action~\cite{Mazharimousavi:2025lld,Myung:2025qqx}
\begin{equation} \label{NED}
\mathcal{L}_{\rm NED}=\frac{1}{16\pi}\Big[2\xi(-\mathcal{F})^{\frac{3}{2}}\Big],\quad \xi=\frac{3\alpha}{2^{3/2}P}
\end{equation}
where the Maxwell term  $\mathcal{F}=F^{\mu\nu}F_{\mu\nu}$ takes $2P^2/r^4$ for a magnetic charge configuration $F_{\theta\varphi}=P\sin \theta$.
In this case, choosing $P=M$ leads to Eq.(\ref{q-em}). However, the selection of $P=M$ is very unnatural.

The scalar field equation is given by
\begin{equation}
\square \phi +\frac{\lambda}{4}f'(\phi) {\cal R}^2_{\rm GB}=0 \label{s-equa}.
\end{equation}
Considering  $G_{\mu\nu}=T^{\rm qOS}_{\mu\nu}$ together  with $\phi=0$ and $f(\phi)=0$,  the qOS-black hole  solution is obtained as
\begin{equation} \label{ansatz}
ds^2_{\rm qOS}= \bar{g}_{\mu\nu}dx^\mu dx^\nu=-g(r)dt^2+\frac{dr^2}{g(r)}+r^2d\Omega^2_2
\end{equation}
whose metric function is given by~\cite{Lewandowski:2022zce}
\begin{equation}
g(r)=1-\frac{2M}{r}+\frac{\alpha M^2}{r^4}, \label{g-sol}
\end{equation}
where the quantum parameter is given by $\alpha=16\sqrt{3}\pi\gamma^3$ with $\gamma$ the dimensionless Barbero-Immirzi parameter. For $\gamma=0.2375$, one finds that $\alpha=1.1663$~\cite{Meissner:2004ju,Domagala:2004jt}.
It is worth noting  that Eq.(\ref{ansatz})  indicates  the qOS-black hole solution without scalar hair.

From $g(r)=0$, one finds two real solutions and two complex solutions
\begin{eqnarray}
&&r_k(M,\alpha),~{\rm for}~k=1,2,3,4, \label{f-roots}
\end{eqnarray}
where $r_{1}$ and $r_{2}$ become complex solutions, while $r_3(M,\alpha)\to r_-(M,\alpha)$ and $r_4(M,\alpha)\to r_+(M,\alpha)$.
The explicit forms of outer/inner horizons are given by
\begin{eqnarray}
r_\pm(M,\alpha)&=&\frac{M}{2}+\frac{1}{2}\Big(M^2+\frac{2^{5/3}M^2\alpha}{(3\eta)^{1/3}}+ \frac{(2\eta)^{1/3}}{3^{2/3}}\Big)^{1/2} \nonumber \\
&\pm& \frac{1}{2}\Big(2M^2-\frac{2^{5/3}M^2\alpha}{(3\eta)^{1/3}}- \frac{(2\eta)^{1/3}}{3^{2/3}}+\frac{2M^3}{(M^2+\frac{2^{5/3}M^2\alpha}{(3\eta)^{1/3}}+ \frac{(2\eta)^{1/3}}{3^{2/3}})^{1/2}}\Big)^{1/2}
 \label{oi-hor}
\end{eqnarray}
with
\begin{equation}
\eta(M,\alpha)=\alpha M^3\Big(9M+\sqrt{3}\sqrt{27M^2-16\alpha}\Big). \label{eta-eq}
\end{equation}
From Eq.(\ref{eta-eq}), one reads off a condition for the existence of  two horizons as
\begin{equation}
0<\alpha<\frac{27M^2}{16},
\end{equation}
which leads to a qOS-extremal black hole for $\alpha=27M^2/16$ as
\begin{equation}
g(r)\to \Big(1-\frac{3M}{2r}\Big)^2\Big(1+\frac{M}{r}+\frac{3M^2}{4r^2}\Big)\to g_e(r,M)\equiv\Big(1-\frac{3M}{2r}\Big)^2.
\end{equation}
In this case, we have the simplest extremal  horizon  from $g_e(r,M)=0$  as
\begin{equation}
r_e(M)=\frac{3M}{2}.
\end{equation}
Using $M_e(\alpha)=\frac{4\sqrt{\alpha}}{3\sqrt{3}}$, one finds
\begin{equation}
g_e(r,\alpha)=\Big(1-\frac{\sqrt{\alpha}}{2\sqrt{3}r}\Big)^2,\quad r_e(\alpha)=\frac{\sqrt{\alpha}}{2\sqrt{3}}.
\end{equation}
Here, we use the mass representation to study the qOS-extremal black hole without loosing  generality.
\begin{figure}
\centering
\includegraphics[angle =0,scale=0.5]{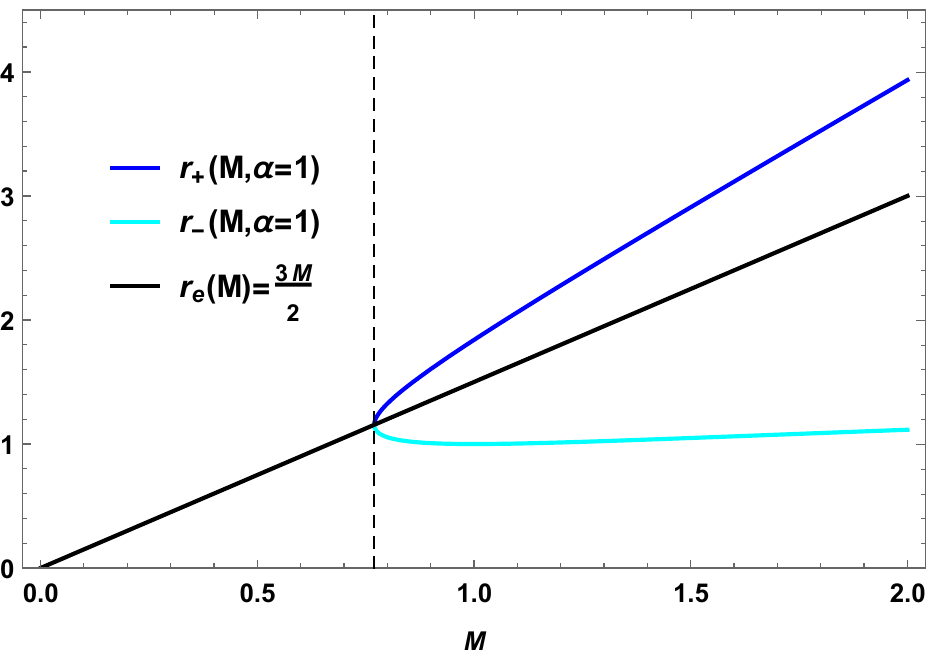}
\caption{ Two  horizons  $r_{\pm}(M,\alpha=1)$ are   function of   $M\in[M_{rem(e)}=0.7698,2]$, showing the lower bound for the mass of black hole.  Here, $r_e(M)=\frac{3M}{2}$ as a function of $M$ represents the extremal horizon, starting from $M>0$.}
\end{figure}
As is shown in Fig. 1, there exist outer/inner horizons  $r_\pm(M,\alpha=1)$ as functions of $M$  with  the lower bound for the mass of  black hole [remnant (extremal) mass $M_{\rm rem(e)}$=0.7698 for $\alpha=1$].
Also, we display the extremal horizon $r_e(M)$ as a function of $M$.

The temperature $T=\frac{\partial m}{\partial S}$ and  heat capacity $C=\frac{\partial m}{\partial r_+}(\frac{\partial T}{\partial r_+})^{-1}$ with mass function $m(M,\alpha)=(r_+^3-r_+^2\sqrt{r_+^2-\alpha})/\alpha$ and area-law enetropy $S=\pi r_+^2$ are given by~\cite{Myung:2025qqx}
\begin{eqnarray}
T(M,\alpha)&=& \frac{2\alpha -3r_+^2(M,\alpha)+3r_+ \sqrt{r_+^2-\alpha}}{ 2 \pi \alpha \sqrt{r_+^2-\alpha}}, \label{tem1} \\
 C(M,\alpha)&\equiv& \frac{NC(M,\alpha)}{DC(M,\alpha)}=-\frac{2\pi r_+(M,\alpha)(r_+^2-\alpha)[2\alpha -3r_+^2+3r_+\sqrt{r_+^2-\alpha}]}{3(\alpha-r_+^2)\sqrt{r_+^2-\alpha}+3r_+^3-4 \alpha r_+}.                         \label{heat1} \\
\end{eqnarray}
Here, the Davies point (blow-up point) can be obtained from solving $DC(M,\alpha)=0$.
We observe from Fig. 2 that  $C(M,1)/|C_S(1,0)|$ blows up at Davies point ($M_D=0.8827$, red dot) where the temperature $T(M_D,1)$ takes  the maximum value. Importantly, we note that  this point coincides with the critical onset mass ($M_c$). The temperature and heat capacity  are zero at remnant (extremal) point ($M_{\rm rem(e)}=0.7698,\bullet$). At this stage, we would like to mention  that the remnant point is equal to the  extremal point.  Their difference is that the remnant point is a starting point for the mass $M$, while the extremal point is the ending point for quantum parameter $\alpha$.
 The qOS black hole is thermodynamically stable if $C>0 (M_e<M<M_D)$, while it is unstable for $C<0(M>M_D)$. Hence, the Davies point is regarded as a critical point which  can represent a sharp phase transition from $C>0$ to $C<0$.

\begin{figure}
\centering
\includegraphics[angle =0,scale=0.5]{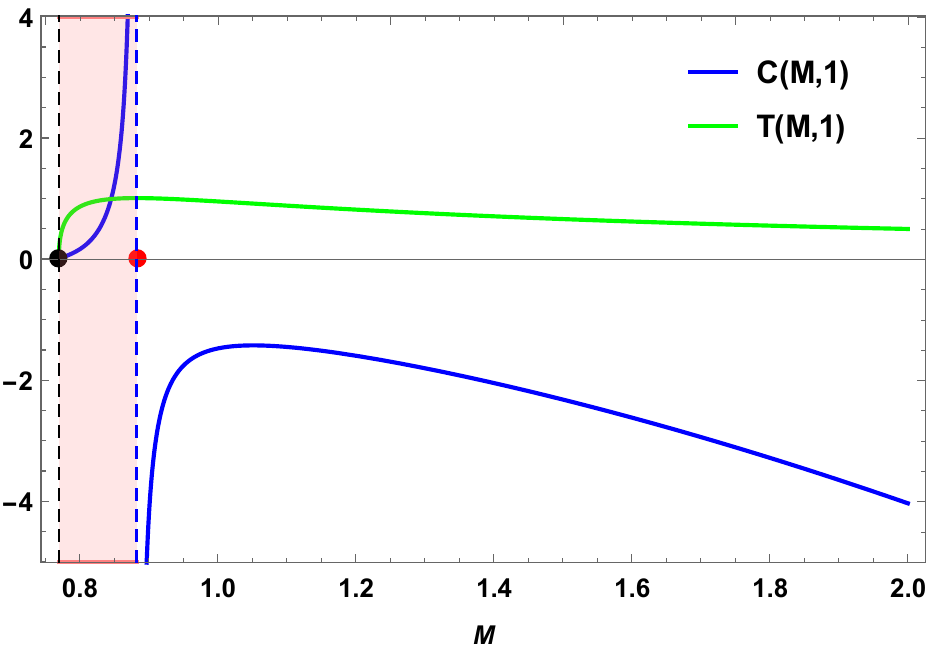}
\caption{ Heat capacity $C(M,\alpha=1)/|C_S(1,0)|$ with $|C_S(1,0)|=25.13$.  Heat capacity blows up at Davies point ($M_D=0.8827,$ red dot) where the temperature $T(M,1)$ has the maximum. This point coincides with the critical onset mass ($M_c$). The heat capacity and temperature are zero at the remnant (extremal) point ($M_{\rm rem(e)}=0.7698,~\bullet$). The shaded region denotes $C>0$, which corresponds to the unstable region of $M_{rem(e)}<M_{th}(\alpha=1,\lambda)\le M_c$ in  Fig. 3(b).  }
\end{figure}

\section{Scalarizations}
\subsection{GB$^-$ scalarization}
In this section, we wish to review briefly   the GB$^-$scalarization with $\lambda<0$.
We introduce the scalar linearized equation
\begin{eqnarray}
  \left(\bar{\square}+ \lambda \bar{{\cal R}}^2_{\rm GB}\right)\delta \phi&=& 0, \label{l-eq2}
\end{eqnarray}
where the overbar  denotes computation based on the qOS-black hole background Eq.(\ref{ansatz}).
Introducing a tortoise coordinate defined by $dr_*=dr/g(r)$ and considering
\begin{equation}
\delta\phi(t,r_*,\theta,\varphi)=\sum_m\sum^\infty_{l=|m|}\frac{\psi_{lm}(t,r_*)}{r}Y_{lm}(\theta,\varphi),
\end{equation}
Eq.(\ref{l-eq2}) reduces  to the Klein-Gordon equation for $s(l=0)$-mode scalar
\begin{equation} \label{mode-d}
\frac{\partial^2\psi_{00}(t,r_*)}{\partial r_*^2} -\frac{\partial^2\psi_{00}(t,r_*)}{\partial t^2}=V(r)\psi_{00}(t,r_*),
\end{equation}
where the potential $V(r)$ is given by
\begin{equation} \label{pot-c}
V(r)=g(r)\Big[\frac{2M}{r^3}-\frac{4\alpha M^2}{r^6}+\tilde{m}^2_{\rm eff}\Big]
\end{equation}
with its effective mass term
\begin{equation}
\tilde{m}^2_{\rm eff}=-\frac{48\lambda M^2}{r^{6}}\Big[\frac{3\alpha^2M^2}{r^6}-\frac{5\alpha M}{r^3} +1\Big].
\end{equation}
For $\lambda>0$ and $\psi_{00}(t,r_*)\sim u(r_*)e^{-i\omega t}$, one found  GB$^+$ scalarization of Schwarzschild black hole with  $\alpha=0$~\cite{Antoniou:2017acq,Doneva:2017bvd,Silva:2017uqg}.
For $\lambda<0$, however, one has obtained quantum parameter ($\alpha$)-mass ($M$)  induced GB$^-$ scalarization for qOS-black holes~\cite{Chen:2025wze,Myung:2025pmx}.

 The onset analysis of spontaneous scalarization  can be analyzed  from  its potential $V(r)$.
To obtain the critical onset parameter, we consider the potential term only
\begin{equation}
V(r)\psi_{00}(t,r_*)=0.
\end{equation}
A critical black hole with $M=M_c$ and $\alpha=\alpha_c$ indicates  the boundary between qOS-black hole and  scalarized qOS-black hole existing  in the limit of $\lambda \to -\infty$.
It could be represented  by  a degenerate  binding potential well whose two turning points
merge at the outer horizon ($r_{\rm out}= r_{\rm in}=r_+$) as
\begin{eqnarray}
 \tilde{m}^2_{\rm eff}\psi_{00}(t,r_*)=0, \quad {\rm for} \quad M=M_c,~\alpha=\alpha_c,~\lambda  \to -\infty.
\end{eqnarray}
In this case, the critical onset mass $M_c$ and quantum parameter $\alpha_c$ are determined by the resonance condition [$rc(M_c,\alpha_c)=0$] because of $\psi_{00}(t,r_*)\not=0$ where the resonance function is defined by
\begin{eqnarray}
rc(M,\alpha)\equiv \frac{3\alpha^2M^2}{r_+^6(M,\alpha)}-\frac{5\alpha M}{r_+^3(M,\alpha)} +1.\label{res-con}
\end{eqnarray}
Here, we check  that $M_c=M_D$ and $\alpha_c=\alpha_D$  by solving  $rc(M,\alpha)=0$  and $DC(M,\alpha)=0$ numerically.
Its coincidence is shown clearly in the Fig. 3(a). This shows a close connection between thermodynamics and scalarization for the qOS black holes.
This implies that the qOS-black holes with $M>M_c$  could not develop the tachyonic instability and it is a forbidden region  for scalarized qOS-black holes.
On the other hand,  the Davies  point is characterized  by the singular behavior of heat capacity at $M=M_D$, which differs  quite from the extremal point $(M=M_e$).
A second order phase transition from $C>0$ to $C<0$ occurs at this  point and this phenomenon is generic for any charged or rotating black holes with two (outer/inner) horizons.
As is shown in Fig. 3(b), the allowed region for GB$^-$ scalaization  is confined to be $M_{rem(e)} \le M_{th}(\alpha=1,\lambda) \le M_c$, which corresponds to the positive regions of  heat capacity ($C>0$).
Here, we expect to find the threshold of instability $ M_{th}(\alpha=1,\lambda)$ numerically  which is an increasing function connecting between  $M=M_{rem(e)}$ and $M=M_c$~\cite{Chen:2025wze}.
On the other hand, the sufficient condition [$M_{sc}(\alpha=1,\lambda)$] for tachyonic instability determined numerically  by $\int^\infty_{r_+(M,\alpha)}\frac{V(r)dr}{g(r)}<0$ is defined as an increasing function within a narrow strip $[M_{rem(e)} \le M_{sc}(\alpha=1,\lambda) \le M_{sc}^{-\lambda\gg \alpha}]$~\cite{Myung:2025pmx}. The sufficiently unstable (shaded)  region is given by $0<M< M_{sc}(\alpha=1,\lambda)$ shown in Fig. 3(b).

\begin{figure*}[t!]
   \mbox{
   (a)
  \includegraphics[width=0.4\textwidth]{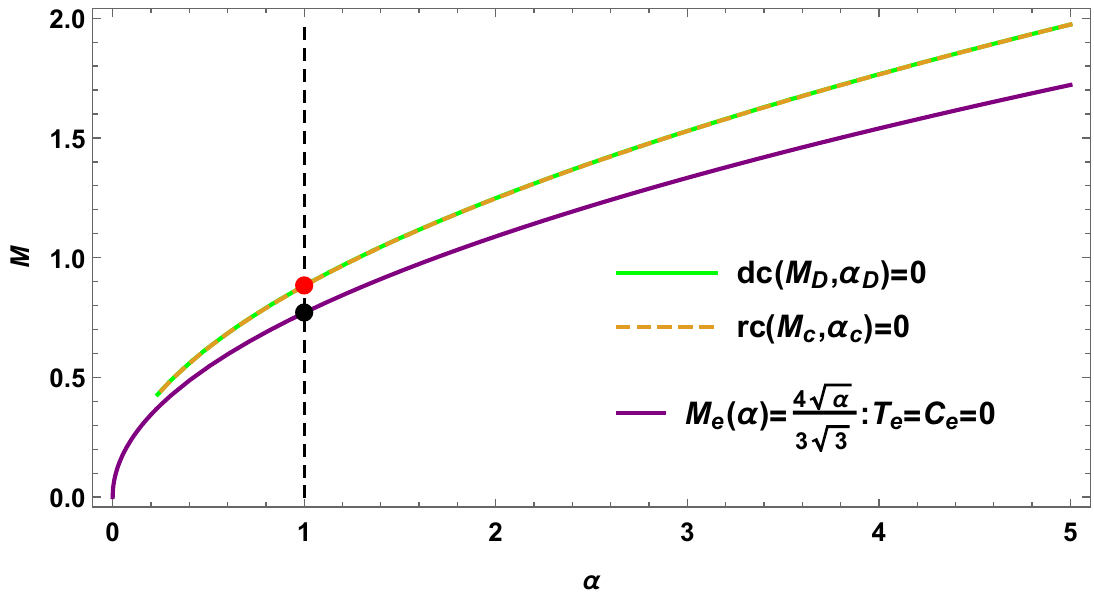}
 (b)
    \includegraphics[width=0.4\textwidth]{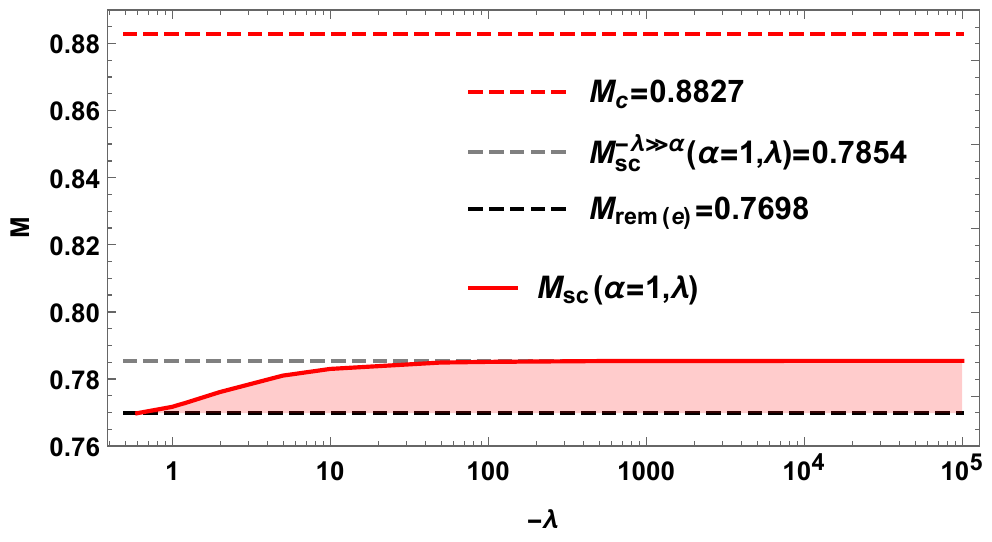}}
\caption{(a) Three curves of $dc(M_D,\alpha_D,0.6)=0$, $rc(M_c,\alpha_c,0.6)=0$, $M_e(\alpha)=\frac{4\sqrt{\alpha}}{3\sqrt{3}}$ for  $\alpha\in[0,5]$ vs  $M\in[0,2]$.
 In general, one finds that $[dc(M_D,\alpha_D,0.6)=0] =[rc(M_c,\alpha_c,0.6)=0]$, implying that the Davies curve is the same as the resonance (critical)  onset curve. However, the extremal curve $M_e(\alpha)$ with $T_e=0$ and $C_e=0$  is not overlapped  from them. For $\alpha=1$, two crossing points denote $M_D=M_c=0.8827$ (red dot) and $M_{rem(e)}=0.7698~(\bullet)$.  (b) Graph for $M_{sc}(\alpha=1,\lambda)$ vs $-\lambda$ for the sufficient condition of instability between $M=M_{sc}^{-\lambda\gg \alpha}=0.7854$ as the upper bound and  $M=M_{rem(e)}$ as the lower bound. The shaded region represents sufficiently unstable region for GB$^-$ scalarization, predicting the single branch of scalarized qOS-black holes.   }
\end{figure*}

\subsection{GB$^{\rm e}$ scalarization for qOS-extremal black hole}
In this section, we wish to focus on the GB$^{\rm e}$ scalarization on the extremal curve $M_e(\alpha)$ with $T_e(\alpha)=C_e(\alpha)=0$ as shown in Fig. 3.
The superscript (e) or subscript (e) denotes extremal but not electric in~\cite{Herdeiro:2021vjo}.
Here, we use $g_e(r,M)$ instead of $g(r,M,\alpha)$, implying that it indicates the single horizon but it is extremal.
In this case, its spacetime is described by
\begin{equation}
ds^2_{\rm e}=-g_e(r,M)dt^2+\frac{dr^2}{g_e(r,M)}+r^2d\Omega^2_2,
\end{equation}
which possesses an AdS$_2 \times S^2$ as the near-horizon geometry.

Here, the radial part of the Klein-Gordon equation  takes the form
\begin{equation} \label{emode-d}
\frac{\partial^2\psi_{00}(t,r_*)}{\partial r_*^2} -\frac{\partial^2\psi_{00}(t,r_*)}{\partial t^2}=V_e(r)\psi_{00}(t,r_*).
\end{equation}
Here,  the $s(l=0)$-mode scalar potential $V_e(r)$ is given by
\begin{equation} \label{epot-c}
V_e(r,M,\lambda)=g_e(r,M)\Big[\frac{3M}{r^3}-\frac{9 M^2}{2r^4}+\tilde{m}^2_e\Big]
\end{equation}
with its effective mass term from $-\lambda \bar{\mathcal{R}}^2_{\rm GB}$
\begin{equation}
\tilde{m}^2_e=-\frac{27\lambda M^2}{2r^{6}}\Big[\frac{15M^2}{r^2}-\frac{24M}{r} +8\Big]. \label{e-masst}
\end{equation}
First of all, we have to mention that the qOS-extremal black hole could not include its critical onset mass because of
$15M^2/r_e^2-24M/r_e+8=-4/3(\not=0)$ for $\lambda\to -\infty$ as is shown in Fig. 3 (no overlapping between extremal and critical onset curves).

For comparison, we introduce the $s$-mode scalar potential of  GB$^+$ scalarization for Schwarzschild BH  with $\lambda>0$ as
\begin{equation} \label{S-pot}
V_S(r,M,\lambda)=g_S(r,M)\Big[\frac{2M}{r^3}-\frac{48\lambda M^2}{r^{6}}\Big],\quad g_S(r,M)=1-\frac{2M}{r}.
\end{equation}
\begin{figure*}[t!]
   \centering
    \mbox{
   (a)
  \includegraphics[width=0.4\textwidth]{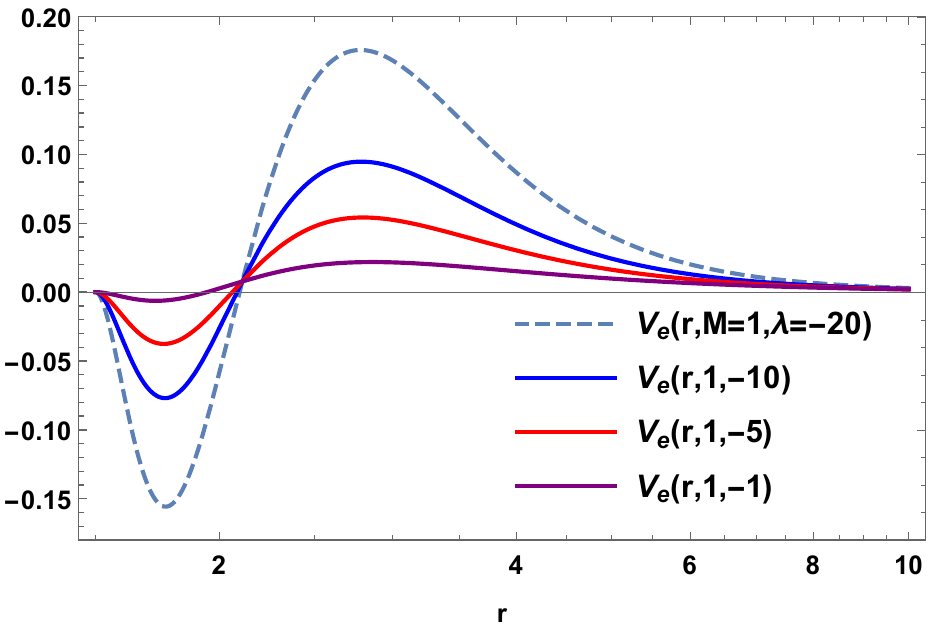}
 (b)
    \includegraphics[width=0.4\textwidth]{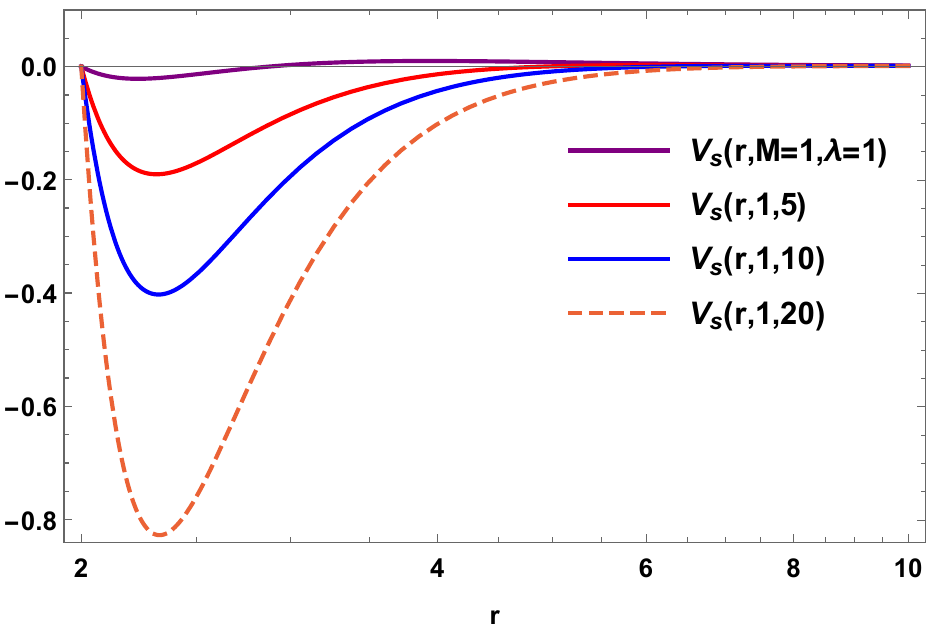}}
\caption{ Scalar potentials.  (a) Extremal scalar potentials $V_e(r,M=1,\lambda)$ with $\lambda=-1,-5,-10,-20$ as functions of $r\in[r_+=1.5,10]$ for GB$^{\rm e}$ scalarization.
(b) Scalar potentials $V_S(r,M=1,\lambda)$ with $\lambda=1,5,10,20$ as functions of $r\in[r_+=2,10]$ for GB$^+$ scalarization. }
\end{figure*}
As is shown in Fig. 4, two have different behaviors: $V_e(r,1,\lambda)$ develops $-$ to $+$ region, while  $V_S(r,1,\lambda)$ develops $-$ region except $\lambda=1$.

To obtain  a sufficient condition for the tachyonic instability, one may use the  condition for instability proposed  by Ref.~\cite{Dotti:2004sh}
\begin{equation}
\int^\infty_{r_e=3M/2}\Big[\frac{V_e(r,M,\lambda)}{g_e(r)}\Big] dr \equiv I_e(M,\lambda)  <0. \label{sc-ta}
\end{equation}
This condition  leads to
\begin{equation}
I_e(M,\lambda)=\frac{70M^2+64\lambda}{315M^3}<0
\end{equation}
which is  solved for $M>0$ with $\lambda<0$ as
\begin{equation}
0<M\le M_{eEEH}(\lambda),\quad M_{eEEH}(\lambda)=\sqrt{\frac{32}{35}} \sqrt{-\lambda} \simeq 0.96 \sqrt{-\lambda}.
\end{equation}
$M_{eEEH}(\lambda)$ is compared to the sufficient condition  $M_{sc}(\alpha=1,\lambda)\in[M_{rem},M_{sc}^{-\lambda\gg\alpha}]$ in Fig. 3(b) for GB$^-$ scalarization.
Here, however,  there are  no upper and lower  bounds on $M$.

On the other hand,  a sufficient condition of  tachyonic instability for GB$^+$ scalarization is given by
\begin{equation}
\int^\infty_{r_+=2M}\Big[\frac{V_S(r,M,\lambda)}{g_S(r,M)}\Big] dr \equiv I_S(M,\lambda)  <0, \label{sc-ta}
\end{equation}
which leads to
\begin{equation}
I_S(M,\lambda)=\frac{5M^2-6\lambda}{20M^3}<0.
\end{equation}
Its inequality is  solved for $M$ with $\lambda>0$ as
\begin{equation}
0<M\le M_{S}(\lambda),\quad M_S(\lambda)=\sqrt{\frac{6}{5}} \sqrt{\lambda} \simeq 1.1 \sqrt{\lambda}.
\end{equation}
We display the sufficiently unstable (shaded) region for the GB$^{\rm e}$ and GB$^+$ scalarization in Fig. 5.
They are similar to each other, but an apparent  difference is $\lambda<0$ for GB$^{\rm e}$ and $\lambda>0$ for GB$^+$. This shows that  GB$^{\rm e}$ scalarization still includes the nature of GB$^-$ scalarization even though its critical onset and remnant (extremal)  points are excluded and thus, there are  no upper and lower bounds on $M$.
\begin{figure*}[t!]
   \centering
    \mbox{
   (a)
  \includegraphics[width=0.4\textwidth]{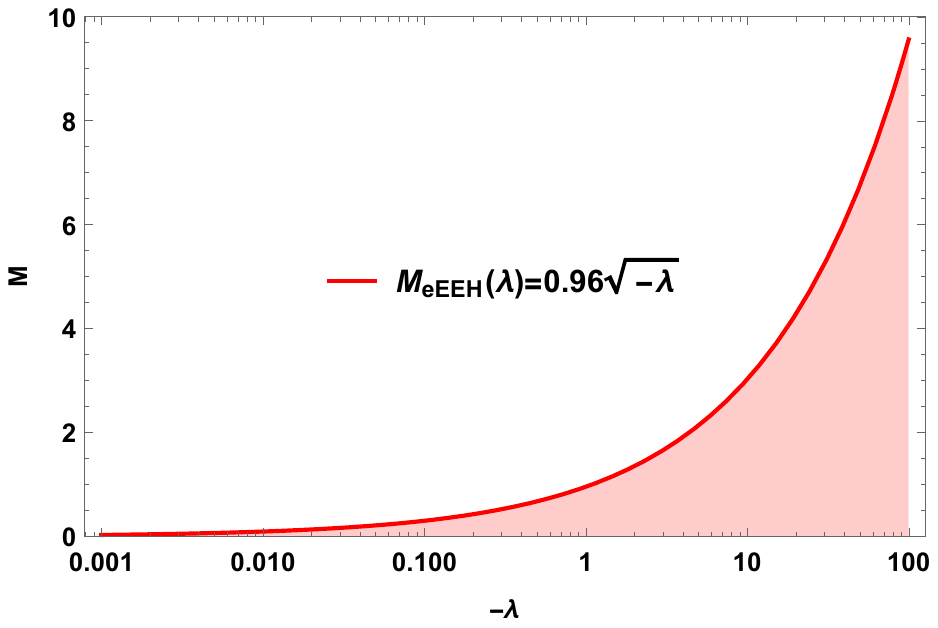}
 (b)
    \includegraphics[width=0.4\textwidth]{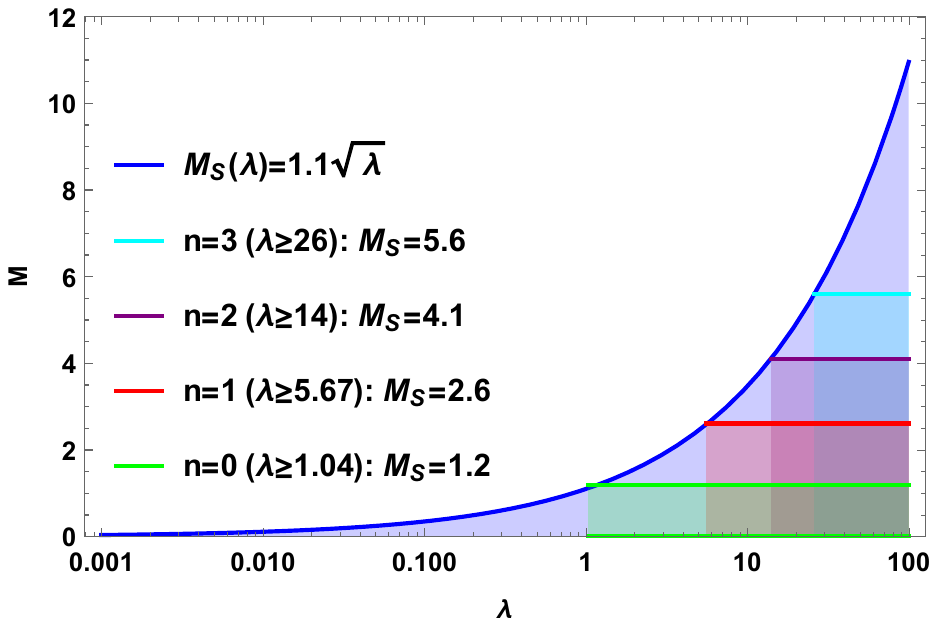}}
\caption{ Two sufficiently unstable (shaded) regions. (a)  qOS-extremal case for GB$^{\rm e}$ scalarization.  A single branch for $\lambda<0$ is allowed for $0<M\le M_{seeH}(=0.96\sqrt{-\lambda})$.
(b) Schwarzschild case  for GB$^+$ scalarization.  Infinite branches starting from  $\lambda_n$ for $n=0,1,2,3\cdots$ are embedded  in  $0<M\le M_{S}(=1.1\sqrt{\lambda})$ for $\lambda>0$.}
\end{figure*}

To explore a further difference on how many branches exist, we use the standard WKB approximation to compute the starting points  $\lambda_n$ with  $n=0,1,2,\cdots$ for existing  branches.
For this purpose, it is necessary to introduce two boundary potentials for qOS-extremal and Schwarzschild black holes as
\begin{eqnarray}
V_{be}^+(r,M)&=&\frac{\sqrt{27} M}{\sqrt{2}r^3}\frac{\sqrt{m^{+2}_{be}(r,M)}}{1-\frac{3M}{2r}},\quad m^{+2}_{be}(r,M)=\frac{15M^2}{r^2}-\frac{24 M}{r} +8, \label{b-pot1} \\
V_{be}^-(r,M)&=&\frac{\sqrt{27} M}{\sqrt{2}r^3}\frac{\sqrt{m^{-2}_{be}(r,M)}}{1-\frac{3M}{2r}},\quad m^{-2}_{be}(r,M)=-\frac{15M^2}{r^2}+\frac{24 M}{r} -8, \label{b-pot1} \\
V_{bS}(r,M)&=& \frac{\sqrt{48} M}{r^3} \frac{1}{\sqrt{1-\frac{2M}{r}}}. \label{b-pot2}
\end{eqnarray}
We plot these potentials in Fig. 6. However, it is observed that $V_{be}^+(r,M=1)$ with $\lambda>0$  is not defined properly for the near-horizon region [$r_+(=1.5)\le r<2.11$] because of  $m^{+2}_{be}(r,1)<0$ for the near-horizon.
Also, $V_{be}^-(r,M=1)$ with $\lambda<0$ is ill-defined  for the far-horizon region of $r>2.11$ because of  $m^{-2}_{be}(r,1)<0$ for the far-horizon.
This implies that  numerical methods cannot solve the Klein-Gordon equation to find out scalar clouds in the extremal black hole background, irrespective of coupling constant $\lambda$~\cite{Richartz:2015saa,Senjaya:2025pyv}.
The numerical investigation is forced  to end at the near-extremal limit~\cite{Hod:2017gvn}.
In addition, this indicates  that a single branch for $\lambda<0$ can exist for $0<M\le M_{sEEH}(=0.96\sqrt{-\lambda})$ and more branches are not attainable for GB$^{\rm e}$ scalarization.
Further, it suggests that   the scalarization in  the near-horizon geometry of AdS$_2\times S^2$ will be explored  separately in the next section.

\begin{figure}
\centering
\mbox{
   (a)
  \includegraphics[width=0.4\textwidth]{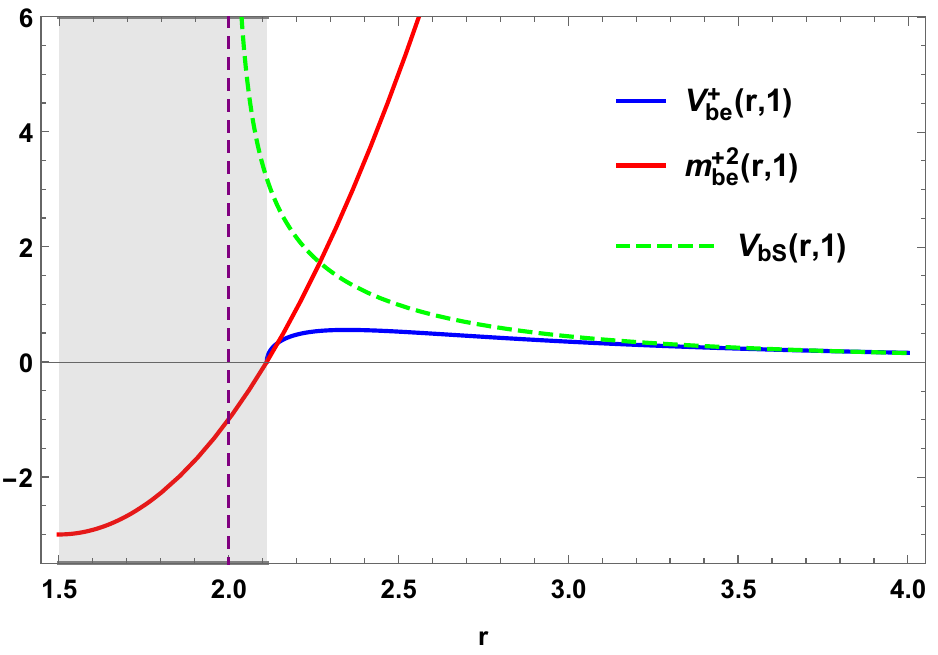}
 (b)
    \includegraphics[width=0.4\textwidth]{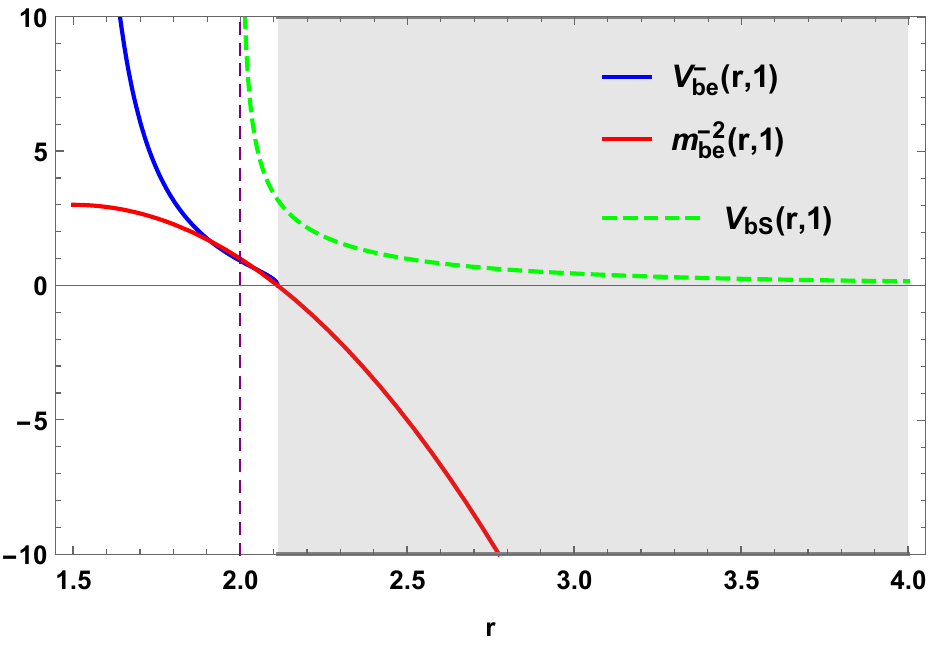}}
\caption{ Two boundary potentials $V_{be}^\pm(r\in[r_+(=1.5),4],M=1)$ for GB$^{\rm e}$ scalarization  and a well-behaved boundary potential  $V_{bS}(r\in[r_+=2({\rm dashed~line}),4],M=1)$ for GB$^+$ scalarization. (a) For $r_+(=1.5)\le r<2.11$ (shaded region located  in the near-horizon region) with $\lambda>0$,  $m^{+2}_{be}(r,1)<0$  implies that the boundary potential  $V_{be}^+(r,M=1)$  is ill-defined because of $V_{be}(r,1)\propto \sqrt{m^{+2}_{be}(r,1)}$. (b) For $ r>2.11$ (shaded region located  in the far-horizon region) with $\lambda<0$,  $m^{-2}_{be}(r,1)<0$  implies that the boundary potential  $V_{be}^-(r,M=1)$  is ill-defined because of $V_{be}^-(r,1)\propto \sqrt{m^{-2}_{be}(r,1)}$. }
\end{figure}

 On the other hand, the GB$^+$ scalarization provides infinite branches whose starting branch points ($\lambda_n$) could be determined by
making use of   the WKB integral
\begin{equation}
\sqrt{\lambda_n}\int^{\infty}_{r_+=2M} V_{bS}(r,M)dr\equiv \sqrt{\lambda_n} I_n(M) =\Big(n+\frac{3}{4}\Big)\pi,\quad  n=0,1,2,\cdots,
\end{equation}
which could be integrated numerically to yield starting branch points as
\begin{equation} \label{alphan}
\lambda_{n}(M)=\frac{\pi^2(n+3/4)^2}{I^2_n(M)},\quad  n=0,1,2,\cdots.
\end{equation}
From this formula, we find four branches whose starting  points are given by  $\lambda_0=1.04,~\lambda_1=5.57,~\lambda_2=14,$ and $\lambda_3=26$.
Accordingly, we embed  the fundamental ($n=0$) branch, the first-excited ($n=1$) branch, $\cdots$ into $M_S(\lambda)$ shown in Fig. 5(b).
However, we could not find a scalar cloud which may  be  a seed for scalarized qOS-extremal black holes existing in the single branch.

\subsection{GB$^{\rm BR}$ scalarization}
In the previous section, we did not obtain  a numerical scalar cloud which is a scalar seed for scalarized qOS-extremal black holes in the single branch.
Here, we wish to find   analytic  scalar clouds which may be  scalar seeds to generate  scalarized qOS-extremal black holes.
For the qOS-extremal black hole, one always finds its near-horizon geometry of the Bertotti-Robinson (BR) background (AdS$_2\times S^2$) as~\cite{deCesare:2024csp}
\begin{equation}
ds_{\rm BR}^2=\Big(\frac{3M}{2}\Big)^2\Big(-\rho^2d\tau^2+\frac{d\rho^2}{\rho^2}\Big) +\Big(\frac{3M}{2}\Big)^2(d\theta^2+\sin^2\theta d\varphi^2), \label{ads-S}
\end{equation}
whose coordinates ($\tau,\rho$) are dimensionless and  the extremal horizon is located at $\rho=0$.
Choosing  $M=2/3$  and inserting Eq.(\ref{ads-S}) into the GB term ($-\lambda\mathcal{R}^2_{\rm GB}\to 8\lambda\to$ mass term $\mu^2$), the $s$-mode linearized equation for $\delta\phi(\tau,\rho)$
is given by
\begin{equation}
-\frac{1}{\rho^2}\partial^2_\tau\delta \phi+\partial_\rho(\rho^2\partial_\rho \delta \phi)-\mu^2 \delta \phi=0. \label{KG-eq}
\end{equation}
Introducing a tortoise coordinate $\rho_*=1/\rho$,
the $s$-mode scalar equation leads to~\cite{Brihaye:2019kvj}
\begin{equation} \label{ads-KGe}
 -\frac{\partial^2\delta \phi(\tau,\rho_*)}{\partial \tau^2}+\frac{\partial^2\delta \phi(\tau,\rho_*)}{\partial \rho_*^2}=V_{\rm GB}(\rho_*,\lambda) \delta \phi(\tau,\rho_*),
\end{equation}
where the GB potential is given by
\begin{equation}
V_{\rm GB}(\rho_*,\lambda)=\frac{\mu^2}{\rho_*^2}\to V_{\rm GB}(\rho,\lambda)=\mu^2 \rho^2.
\end{equation}
\begin{figure}
\centering
\mbox{
   (a)
  \includegraphics[width=0.4\textwidth]{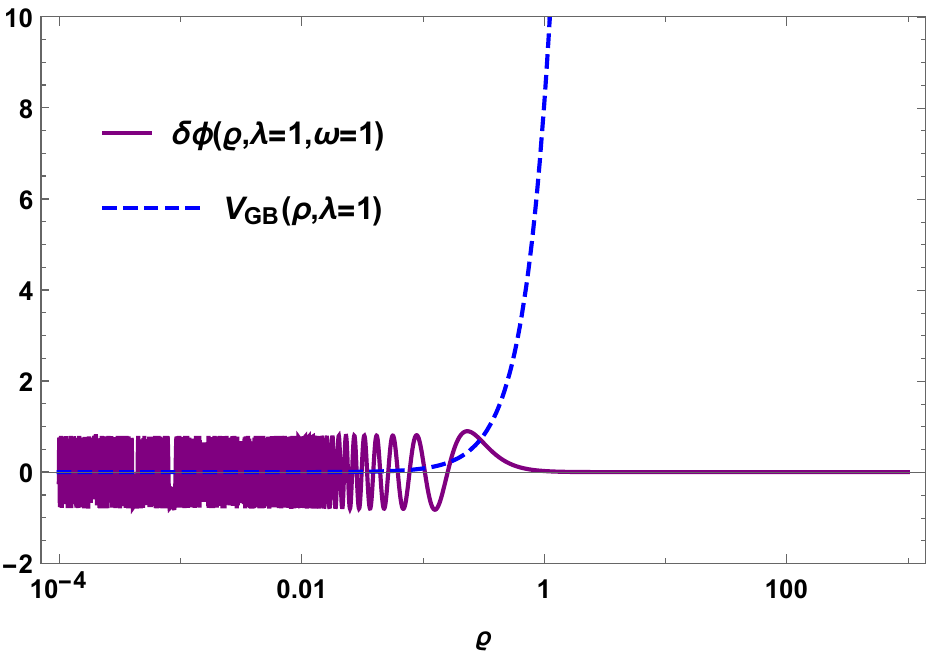}
 (b)
    \includegraphics[width=0.4\textwidth]{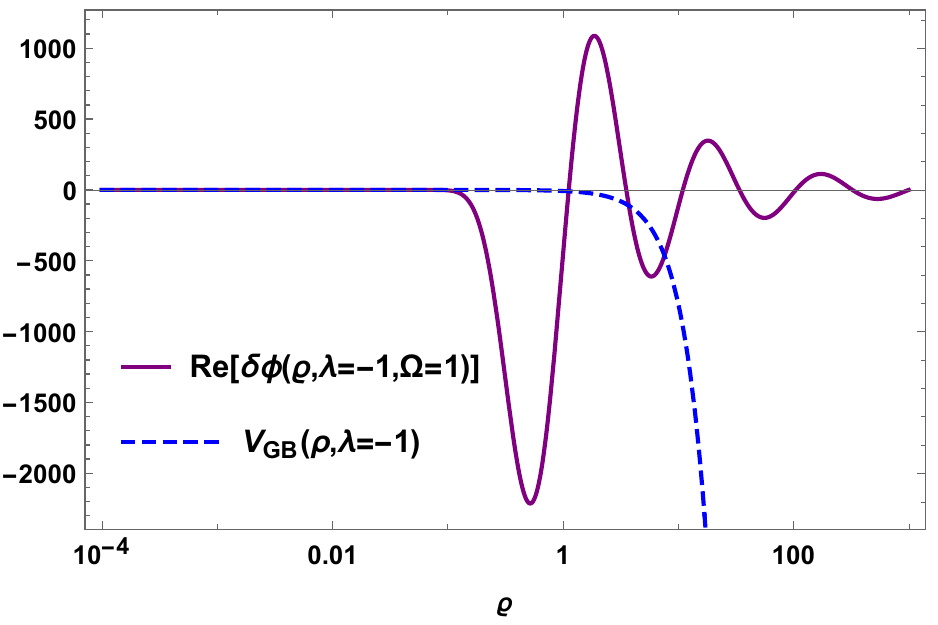}}
\caption{ Two scalar solutions. (a) Regular scalar solution of $\delta\phi(\rho,\lambda=1)$  with energy $\omega^2=1$ and its positive potential $V_{\rm GB}(\rho,\lambda=1)=8\rho^2$.
 (b) Tachyonic scalar solution   Re[$\delta \phi(\rho,\lambda=-1)]$ with $\Omega^2=1$ and its negative potential  $V_{\rm GB}(\rho,\lambda=-1)=-8\rho^2$.  }
\end{figure}
At this stage, we introduce the Breitenlohner-Freedman bound for a massive scalar propagating around the AdS$_2$ spacetime~\cite{Breitenlohner:1982jf,Breitenlohner:1982bm}
\begin{equation}
\mu^2\ge \mu^2_{BF}=-\frac{1}{4},
\end{equation}
whose solution below it corresponds to tachyons in AdS$_2$ spacetime and this AdS$_2$ becomes unstable.
 \begin{figure}
\centering
\mbox{
   (a)
  \includegraphics[width=0.4\textwidth]{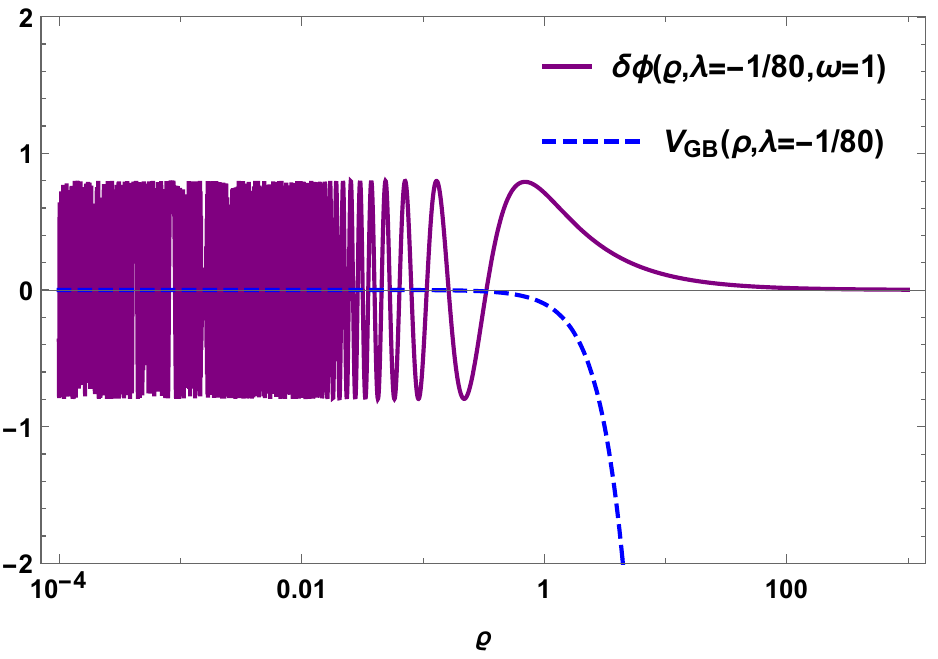}
 (b)
    \includegraphics[width=0.4\textwidth]{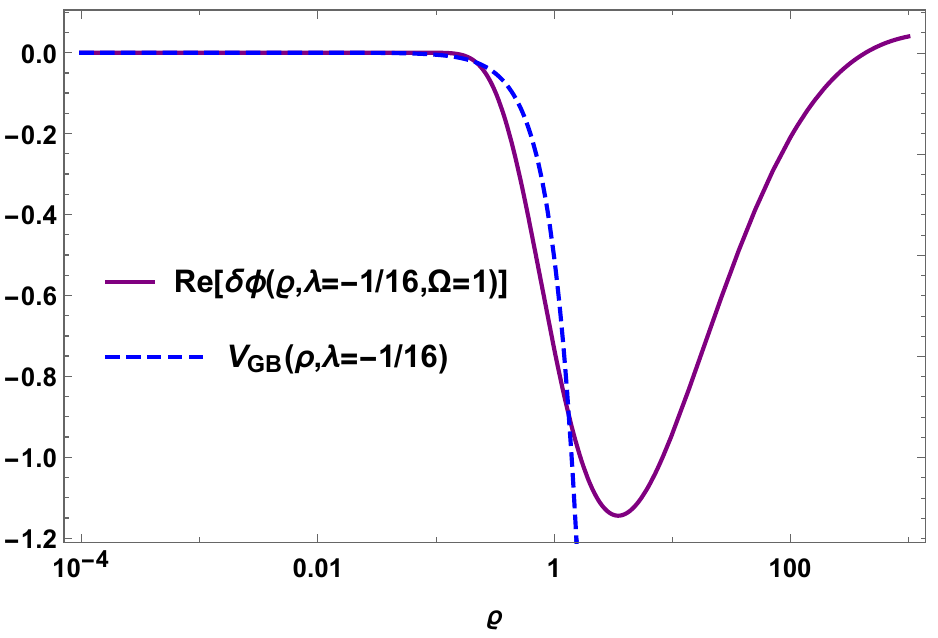}}
\caption{ Two scalar solutions with negative mass. (a) Regular scalar solution of $\delta\phi(\rho,\lambda=-1/80)$  with energy $\omega^2=1$ and its negative potential $V_{\rm GB}(\rho,\lambda=-1/80)=-0.1\rho^2$. This case has mass $\mu^2=-0.1>\mu^2_{BF}$.
 (b) Tachyonic scalar solution   Re[$\delta \phi(\rho,\lambda=-1/16)]$ with $\Omega^2=1$ and its negative potential  $V_{\rm GB}(\rho,\lambda=-1/16)=-0.5\rho^2$. Its mass is given by $\mu^2=-0.5<\mu^2_{BF}$.  }
\end{figure}
Considering  $\delta \phi(\tau,\rho_*)=e^{-i\omega \tau} \delta \phi(\rho_*)$, Eq.(\ref{ads-KGe}) takes the Schr\"odinger-type equation
\begin{equation} \label{ads-Sch}
\frac{\partial^2\delta \phi(\rho_*)}{\partial \rho_*^2}+\Big[\omega^2-V_{\rm GB}(\rho_*,\lambda)\Big] \delta \phi(\rho_*)=0
\end{equation}
whose normalizable solution is given  by the first-kind Bessel function with standard mass $\mu^2=8\lambda>-1/4$  as~\cite{Myung:2011uy}
\begin{equation}
\delta \phi(\rho_*)= \sqrt{\rho_*}J_\nu(\omega\rho_*) \to \delta \phi(\rho)= \frac{1}{\sqrt{\rho}}J_\nu\Big(\frac{\omega}{\rho}\Big), \quad \nu=\frac{\sqrt{32\lambda +1}}{2}. \label{exact-sol1}
\end{equation}
Here, we note that the event horizon is located at $\rho_* \to \infty~(\rho\to 0)$, while the infinity is located at $\rho_*\to 0(\rho\to \infty)$ [see Fig. 7(a) for $\lambda=1$].
Actually, this corresponds to a regular scalar solution with infinite nodes because $\delta \phi(\rho)$ is finite on the horizon and it approaches zero at infinity. This is surely  a stable solution propagating around AdS$_2$ spacetime.

On the other hand, considering  $\delta \phi(\tau,\rho_*)=e^{\Omega \tau} \delta \phi(\rho_*)$  with an exponentially growing mode with $\tau$,
 Eq.(\ref{ads-KGe}) takes the form
\begin{equation} \label{ads-Sch}
\frac{\partial^2\delta \phi(\rho_*)}{\partial \rho_*^2}-\Big[\Omega^2+V_{\rm GB}(\rho_*,\lambda)\Big] \delta \phi(\rho_*)=0,
\end{equation}
whose tachyonic  solution is given  by the second-kind Bessel function  with tachyonic mass $\mu^2=8\lambda<-1/4$
\begin{equation}
\delta \phi(\rho_*)=\sqrt{\rho_*}Y_\nu(i\Omega\rho_*) \to \delta \phi(\rho)=\frac{1}{\sqrt{\rho}}Y_\nu\Big(\frac{i\Omega}{\rho}\Big),\quad \nu=\frac{\sqrt{32\lambda +1}}{2}. \label{exact-sol2}
\end{equation}
Its real part  is depicted  in Fig. 7(b) with $\lambda=-1$. It seems  not to be  a normalizable  solution because it takes a large value of $-2000$ (a large pulse) even though it takes zero at the horizon and infinity.
This corresponds to an unstable solution. In Fig. 8, one checks the BF bound that  a regular solution is allowed for  $\mu^2=-0.1>\mu^2_{BF}$ and a finite tachyonic solution appears with $\mu^2=-0.5<\mu^2_{BF}$.
\begin{figure}
\centering
\mbox{
   (a)
  \includegraphics[width=0.4\textwidth]{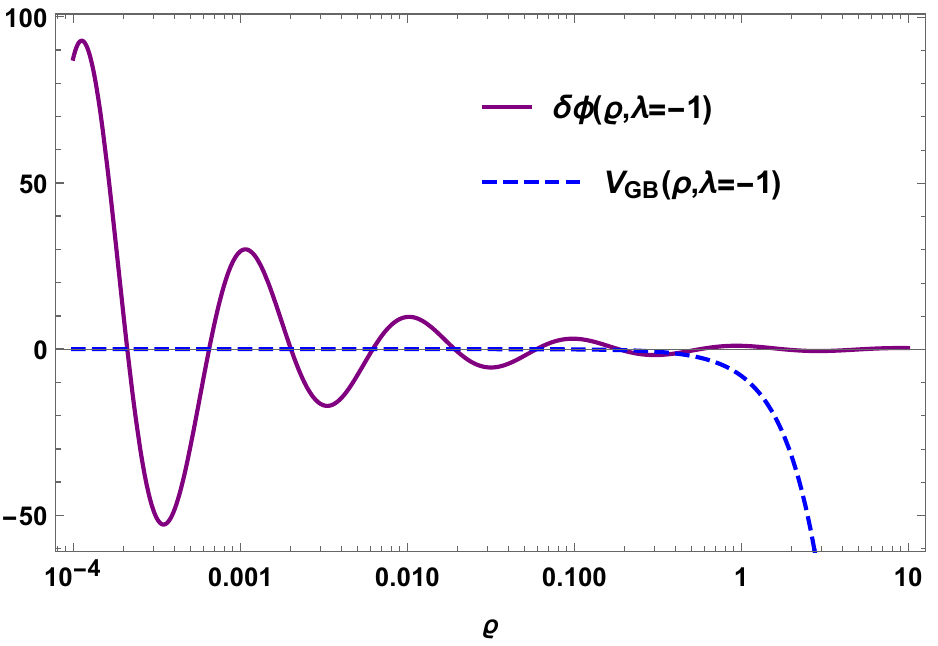}
 (b)
    \includegraphics[width=0.4\textwidth]{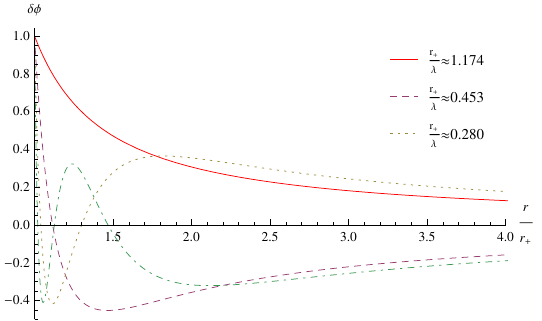}}
\caption{ Two different scalar clouds. (a) Tachyonic (large) scalar cloud of $\delta \phi(\rho,\lambda=-1)$ and its negative  potential $V_{\rm  GB}=-8\rho^2$ with zero energy ($\omega^2=0$) for GB$^{\rm BR}$ scalarization.
 This has  many nodes.
  (b) Regular scalar clouds  $\delta \phi_n(r,r_+=2)$ with $n=0,1,2$ for GB$^+$ scalarization~\cite{Myung:2018iyq}. Here, $n$ represents number of nodes (number of zero-crossings at $r$-axis).  }
\end{figure}

Importantly, solving the static scalar equation for  $\omega=0$ whose time-dependence is nothing as
\begin{equation}
\frac{\partial^2\delta \phi(\rho_*)}{\partial \rho_*^2} -V_{\rm GB}(\rho_*,\lambda)\delta \phi(\rho_*)=0,
\end{equation}
one finds a scalar cloud for the single branch
\begin{equation}
\delta \phi(\rho_*,\lambda)=c_1(\rho_*)^{\frac{1}{2}+\nu}+c_2(\rho_*)^{\frac{1}{2}-\nu} \to\delta \phi(\rho,\lambda)=c_1(\rho)^{-\nu-\frac{1}{2}}+c_2(\rho)^{\nu-\frac{1}{2}}. \label{scl-eq}
\end{equation}
Choosing $\lambda=-1$ and $c_1=c_2=1/2$, the tachyonic  seed and its potential are given by
\begin{equation}
\delta \phi(\rho,\lambda=-1)=\frac{1}{\sqrt{\rho}}\cos\Big[\frac{\sqrt{31} \ln(\rho)}{2}\Big], \quad V_{\rm GB}(\rho,\lambda=-1)=-8  \rho^2, \label{tach-s}
\end{equation}
which  has many nodes as is shown Fig. 9(a) but it takes a large value of 100 at $\rho=10^{-4}$. This tachyonic cloud is considered as a new feature to represent  onset scalarization of qOS-extremal black holes.

On the other hand, one finds regular (finite at the horizon)  scalar clouds labelled by  number of nodes ($n=0,1,2,\cdots$) for GB$^+$ scalarization for Schwarzschild black holes [see Fig. 9(b)].
These  were obtained  by numerical computations~\cite{Myung:2018iyq}:
 $\delta \phi_0(r,M=1)$ has zero node (zero crossing at $r$-axis) with $\lambda_0=0.73$, $\delta \phi_1(r,M=1)$ has one node with $\lambda_1=4.87$,  and $\delta \phi_2(r,M=1)$ has two nodes with $\lambda_2=12.8$. We note that these starting branch points ($\lambda_n$) are slightly different from those in Fig. 5(b) predicted  by the WKB approximation.

 Finally, fixing $\lambda=1$, $c_1=1$, and  $c_2=0$, one finds from  Eq.(\ref{scl-eq}) as
 \begin{equation}
 \delta_{inf} \phi(\rho,\lambda=1)=\rho^{-\frac{1}{2}(\sqrt{33}+1)}, \label{inf-sc}
 \end{equation}
 which shows that $\delta_{inf} \phi(\rho,\lambda=1)$  approaches infinity as $\rho\to 0$ but it is zero at $\rho=\infty$ (see Fig. 10) and thus, it is called the blow-up scalar cloud at the horizon.
  We note that the other term of $\rho^{\frac{1}{2}(\sqrt{33}-1)}$ approaches zero as $\rho\to 0$, while it takes the infinity as $\rho \to \infty$, corresponding to a non-normalizable solution.

\begin{figure}
\centering
  \includegraphics[width=0.5\textwidth]{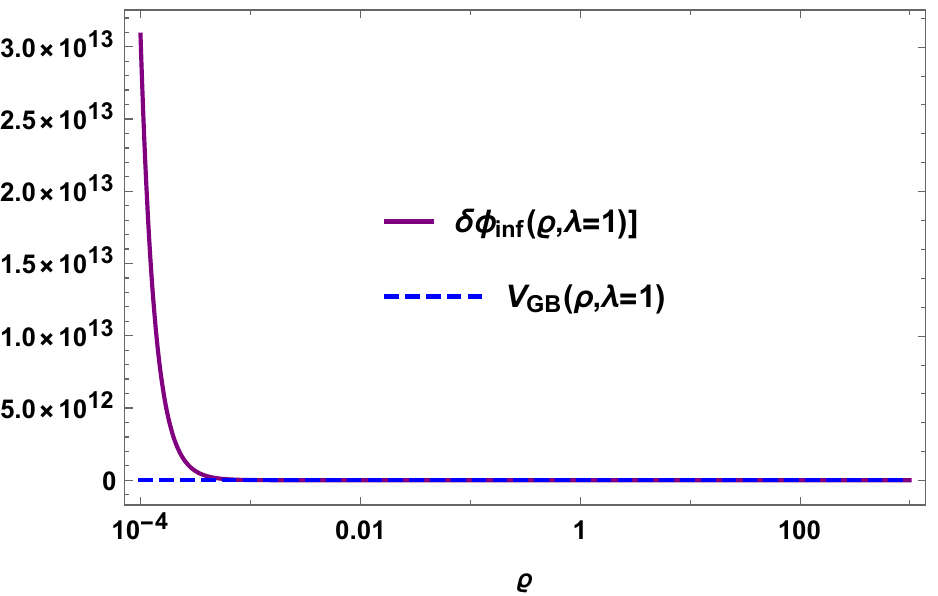}
\caption{ Infinite scalar cloud of $\delta \phi_{inf}(\rho,\lambda=1)$ as $\rho\to \infty$  and its positive  potential $V_{\rm  GB}=8\rho^2$ with $\lambda=1$.
 This blow-up at $\rho=0$  might be  related to the Aretakis instability.   }
\end{figure}

\section{Aretakis instability}
In the previous section, we found that the tachyonic cloud may take the large value at the horizon of $\rho=0$ and the scalar cloud possesses the infinity (blow-up) at the horizon, suggesting other instability.
There were no such  large  scalar clouds for known onset scalarizations of non-extremal black holes because scalar clouds play the role of seeds to generate infinite branches of scalarized black holes.
Hence, we have to identify  their  nature of tachyonic and infinite scalar clouds: onset scalarization of extremal black holes~\cite{Brihaye:2019kvj} or other instability.
To make a connection to  other instability, we may consider the Aretakis instability (classical linear instability)~\cite{Aretakis:2011ha,Katagiri:2021xig,Chen:2025sim}, which captures a feature of any propagating scalar with standard mass around extremal black holes.

In order to study the Aretakis instability, we introduce an ingoing time coordinate $v=\tau-1/\rho$ with $M=2/3$.
Then,  the near-horizon geometry can be described by  ingoing Eddington-Finkelstein coordinates $(v,\rho,\theta,\varphi$) as
\begin{align}
	ds^2_{EF}=-\rho^2dv^2+2dvd\rho+d\theta^2+\sin^2\theta d\varphi^2.
\end{align}
The linearized equation for $s(l=0)$-mode $\delta \phi(v,\rho)$ takes the form
\begin{align}
	2\partial_v\partial_\rho\delta\phi+\partial_\rho\left(\rho^2\partial_\rho\delta\phi\right)-\mu^2\delta\phi=0,\quad \mu^2=8\lambda,
\end{align}
where the first term differs from that of Eq.(\ref{KG-eq}). Hence, we note that  its time-independent equation is the same as in  Eq.(\ref{KG-eq}).

Acting the operator $\partial_\rho^N$ to the above equation and evaluating it at the horizon of $\rho=0$, one finds
\begin{align}
	2\partial_v\partial_\rho^{N+1}\delta\phi =[8\lambda-N(N+1)]\partial_\rho^N\delta\phi.
\end{align}
The Aretakis constant can be defined
\begin{align}
	H_N=\partial_\rho^{N+1}\delta\phi,
\end{align}
only if
\begin{align}
	N(N+1)=8\lambda.
\end{align}
This can be  solved   for a positive integer $N$ as
 \begin{equation}
 N=\nu-\frac{1}{2}, \quad {\rm for} \quad  \nu=\frac{\sqrt{32\lambda +1}}{2}
 \end{equation}
which implies $\lambda>0$ (standard mass term). This means that the Aretakis constant (horizon hair) has nothing to do with the tachyonic scalar cloud which takes the large value at $\rho=0$.
The late-time behavior in the near-horizon region  takes the form  when using operator method to solve  the lowest-weight condition of $L_-\delta \phi_{N,h}=0$ with $L_-=v^2\partial_v-2(\rho v+1)\partial _\rho$ and the lowest-weight $h= N+1$~\cite{Chen:2025sim}
\begin{align}
	\delta\phi_{N,N+1}(v,\rho)\propto v^{-N-1}(v \rho +2)^{-N-1}, \label{phi-vac}
\end{align}
which corresponds to Eq.(\ref{inf-sc}) for $\rho$-dependence with the AdS scaling dimension $\Delta=N+1=\nu+1/2$.
In this case, the higher weight elements  of $\delta\phi_{N,n+N+1}$ can be generated by $n$-repeated actions $(L_+)^n=\partial^n_v$
\begin{equation}
\delta\phi_{N,n+N+1}=(L_+)^n\delta\phi_{N,N+1} \propto v^{-n-N-1}.
\end{equation}
Furthermore, one obtains from Eq.(\ref{phi-vac})
\begin{align}
	\partial_\rho^{k}\delta\phi_{N,N+1}|_{\rho\to 0} \propto v^{k-N-1},
\end{align}
which  implies  that $\partial_\rho^{k\leq N}\delta\phi_{N,N+1}|_{\rho\to 0}$ decays at late times, whereas $\partial_\rho^{k= N+1}\delta\phi_{N,N+1}|_{\rho\to 0}$ is a constant $H_N$.
This becomes the Aretakis instability  if the coupling constant $\lambda$ and its mass $\mu^2$ are positive ($N$: positive integer) with $k\ge N+2$ because $\partial_\rho^{k\ge N+2}\delta\phi_{N,N+1}|_{\rho\to 0}$ grows polynomially in the ingoing time $v$ with a power of $k-N-1$.
For $\lambda=1/4(N=1,\mu^2=2)$, we have $\partial_\rho \delta\phi_{1,2}|_{\rho\to 0}\propto1/v,~\partial^2_\rho \delta\phi_{1,2}|_{\rho\to 0}\propto 1(=H_1),~\partial^3_\rho \delta\phi_{1,2}|_{\rho\to 0}\propto v$.
This case is  related to the infinite scalar cloud given by Eq.(\ref{inf-sc}) found for $\lambda >0$.
However, it is clear  that the tachyonic scalar cloud  Eq.(\ref{tach-s}) has nothing to do the Aretakis instability  because it  was found for $\lambda <0$.
Hence, it is reasonable to say that the appearance of the large scalar cloud at the horizon ($\rho=0$) is a new  feature to represent  onset scalarization for extremal black holes via tachyon  with a negative mass $\mu^2=8\lambda<0$~\cite{Brihaye:2019kvj}.

\section{Discussions}

First of all, we would like to mention the thermodynamics and GB$^-$ scalarization for  qOS-(non-extremal) black holes described by mass ($M$) and quantum parameter ($\alpha$) found in the EGBS theory.
There was  a strong connection ($M_D=M_c$) between thermodynamics ($M_D$: Davies point) and GB$^-$ scalarization
($M_c$: critical onset mass) for the qOS black holes~\cite{Myung:2025pmx}.
This implies that the qOS-black holes with $M>M_c$  could not develop the tachyonic instability and it corresponds to a forbidden region  for scalarized qOS-black holes.
The allowed region  for GB$^-$ scalarization  is given by a narrow region of $M_{rem(e)}(=0.7698)<M\le M_c(=0.8827)$ with quantum parameter $\alpha=1$, which corresponds to positive heat capacity (thermodynamically stable region).

In the present work, we have investigated  scalarization of qOS-extremal black holes described by mass ($M$) in the EGBS theory with the unknown action $\mathcal{L}_{\rm qOS}$.
Here, the quantum parameter $\alpha$  is redundant because of the extremal condition ($\alpha=27M^2/16$).  Also, its temperature and heat capacity were always  zero and critical onset parameter $M_c$ disappeared.
Focusing on the onset of GB$^-$ scalarization with $\lambda<0$, we found  the sufficiently unstable region of $0<M\le M_{sEEH}(=0.96\sqrt{-\lambda})$. This  predicts the appearance of the single branch of scalarized qOS-extremal black holes. Interestingly,  this could be   compared to the sufficiently unstable region of $0<M\le M_{S}(=1.1\sqrt{\lambda})$ for GB$^+$ scalarization of Schwarzschild black holes which embeds infinite branches ($n=0,1,2,\cdots$) of scalarized black holes.  However, we could not obtain its tachyonic scalar cloud which  may be  a seed to generate the single branch of scalarized qOS-extremal black holes.
This is because  numerical methods (for example, WKB approximation)  cannot be used to solve the Klein-Gordon equation to find out scalar clouds in the extremal black hole background~\cite{Richartz:2015saa,Senjaya:2025pyv}.
This forces the numerical investigation to end at the near-extremal limit~\cite{Hod:2017gvn}.

To obtain the tachyonic scalar cloud with tachyonic mass $\mu^2=8\lambda<0$, we have considered   the near-horizon geometry of the Bertotti-Bobinson (AdS$_2\times S^2$) spacetime.
In this case, we found  the appearance of  a large scalar cloud [Eq.(\ref{tach-s}) and Fig. 9(a))] at the horizon ($\rho=0$). This  is surely the new feature to represent  onset scalarization of  extremal black holes for the tachyon with negative mass $\mu^2=8\lambda<0$.
However, it is not related to the Aretakis instability of a propagating scalar with standard mass $\mu^2=8\lambda>0$  around the AdS$_2\times S^2$ spacetime.
This instability indicates  polynomial instability of the ingoing time $v$ at the horizon of $\rho\to 0$. Also, the Aretakis instability is  related to the static scalar infinity  at $\rho=0$ [Eq.(\ref{inf-sc}) and Fig.10] with positive mass $\mu^2=8\lambda>0$. This static scalar infinity might  not be  considered as a proper scalar cloud to generate scalarized qOS-extremal black holes.

Finally, we have a restriction on constructing scalarized qOS-extremal black holes because of the unknown ${\cal L}_{\rm qOS}$.
For this purpose, it would be better to construct scalarized qOS-extremal black holes if one knows  ${\cal L}_{\rm qOS}$.

\section{Acknowledgments}
The author thanks Hong Guo for helpful discussions on Aretakis instability.
The author  is supported by the National Research Foundation of Korea (NRF) grant
 funded by the Korea government (MSIT) (RS-2022-NR069013).

\end{document}